\documentclass[aps,pra,twocolumn,amsmath,amssymb, superscriptaddress,tightenlines]{revtex4}
\usepackage{graphicx}
\usepackage{epsfig}
\usepackage{makecell}
\usepackage{dcolumn}
\usepackage{bm}
\usepackage{amssymb}
\usepackage{bm}
\usepackage{amsmath, bm}
\usepackage{upgreek}
\usepackage{enumitem}  
\usepackage[colorlinks,citecolor=blue,linkcolor=blue,hyperindex]{hyperref}

\begin{document}

\title{Competing phases and intertwined orders in coupled wires near the self-dual point}
\author{Ken K. W. Ma}
\affiliation{National High Magnetic Field Laboratory, Tallahassee, Florida 32310, USA}
\author{O\u{g}uz T\"urker}
\affiliation{National High Magnetic Field Laboratory, Tallahassee, Florida 32310, USA}
\author{Alexander Seidel}
\affiliation{Technical University of Munich, TUM School of Natural Sciences, Physics Department, 85748 Garching, Germany}
\affiliation{Munich Center for Quantum Science and Technology (MCQST), Schellingstr. 4, 80799 M{\"u}nchen, Germany}
\affiliation{Department of Physics, Washington University, St.
Louis, Missouri 63130, USA}
\author{Kun Yang}
\affiliation{Department of Physics, Florida State University, Tallahassee, Florida 32306, USA}
\affiliation{National High Magnetic Field Laboratory, Tallahassee, Florida 32310, USA}

\date{\today}


\begin{abstract}

The interplay between different quantum phases plays an important role in strongly correlated systems, such as high-$T_c$ cuprates, quantum spin systems, and ultracold atoms. In particular, the application of effective field theory model and renormalization group analysis suggested that the coexistence of density wave (DW) and superfluid (SF) orders can lead to a supersolid phase of ultracold bosons. Here we revisit the problem by considering weakly coupled wires, where we treat the intra-wire interactions exactly via bosonization and inter-wire couplings using a mean-field theory which becomes asymptotically exact in the limit of high dimensionality. We obtain and solve the mean-field equations for the system near the self-dual point, where each wire has the  Luttinger parameter $K=1$ and the inter-wire DW and SF coupling strengths are identical. This allows us to find explicit solutions for the possible supersolid order. An energy comparison between different possible solutions shows that the supersolid order is energetically unfavorable at zero temperature. This suggests that the density wave and superfluid phases are connected by a first order transition near the self-dual point. We also discuss the relation between our work and the intertwining of charge density wave and superconducting orders in cuprates.

\end{abstract}

\maketitle

\section{Introduction}  \label{sec:intro}

Since the late 1960s, supersolid has become one of the long-sought phases in quantum materials~\cite{Andreev1969}. It was originally believed that helium-4 would be a promising platform to observe supersolidity~\cite{Nepomnyashchii1971, Schneider1971}. Some experimental evidence had indeed been reported~\cite{Kim2004-1, Kim2004-2}. However, the interpretation there was seriously questioned~\cite{Svistunov2005, Ye2006, Day2007, Prokofev2007, Ceperley, Anderson1, Anderson2, Anderson3, Hunt2009, Balibar2010, Kim2010, Boninsegni2012, Kim2012}. The idea of supersolidity has attracted extensive studies because its realization requires breaking of translational and U(1) symmetries simultaneously. On one hand, this indicates the coexistence of crystalline and superfluid orders~\cite{Chester1970, Leggett1970}. On the other hand, this coexistence seems to contradict the common intuition that different orders usually compete and suppress each other. 

Thanks to the discovery of high-$T_c$ superconducting cuprates and related studies on quantum spin systems, researchers have gained a better understanding of the above ``dilemma". It is common for those strongly correlated materials to have complex phase diagrams, so that there is an interplay between different broken-symmetry phases~\cite{Emery1998}. Instead of always competing against and suppressing each other, previous work pointed out that different orders can actually coexist. This led to the concept of intertwined orders~\cite{RMP2015}. Meanwhile, the breakthrough in cold atom experiment has provided a more controllable platform for studying strongly correlated systems. Different quantum phases that were originally introduced for electronic and magnetic systems have been realized in cold atoms~\cite{Bloch2002}. Taking advantage of this development, various cold atom systems for realizing supersolid have been proposed~\cite{Pohl2010, Shlyapnikov2015, Pupillo2010, Boninsegni-Low, Pfau2021}. While most of the existing theories have focused on supersolid that breaks \textit{discrete} translational symmetry, convincing evidence for supersolidity that actually breaks \textit{continuous} translational symmetry was reported recently in ultracold bosons~\cite{Donner2017, Ketterle2017, dipolar-SS1, dipolar-SS2, dipolar-SS3, dipolar-SS4}. It is thus worthwhile to explore theoretically under what conditions a supersolid that breaks continuous translation symmetry can be realized.

Motivated by the above discussion, we revisit a particular proposal of realizing supersolid for bosons trapped in an array of one-dimensional tubes~\cite{Carr}. The system may exhibit a density wave (DW) or a pure superfluid (SF) order, whereas their coexistence corresponds to a supersolid order. This scenario resembles the interplay between charge density wave (CDW) and superconducting (SC) orders in a high-$T_c$ superconducting cuprate, which was modeled as an array of coupled one-dimensional stripes or chains~\cite{Tsvelik2002, Fradkin2010}. Originally, an interchain mean field treatment at $K=1$ led to the conclusion that the CDW and SC phases are connected by first order transition. There is no coexistence of the two orders except at this self-dual point~\cite{Tsvelik2002} (the meaning of this self-duality will be elucidated later). In the (2+1)-dimensional regime, an effective nonlinear sigma model with an enlarged O(4) symmetry was formulated at the point $K=1$~\cite{Fradkin2010}. Based on renormalization group (RG) analysis, it was argued that the coexistence of CDW and SC orders is possible. Meanwhile, no detailed analysis was performed when the system is away from the self-dual point, apart from claiming that the corresponding symmetry breaking term in the nonlinear sigma model is marginally irrelevant~\cite{Fradkin2010, Carr}.

Here, we revisit this problem using a different yet more quantitative approach. After a mean field treatment of inter-wire couplings, the system reduces to a collection of decoupled wires. Each of them is described by a double sine-Gordon model. This model cannot be solved analytically in general. Meanwhile, it is possible to combine the two cosine terms into a single cosine term at $K=1$~\cite{Levin, Nersesyan}. The resulting standard sine-Gordon model is integrable. In particular, its soliton mass and ground state energy density have been determined analytically~\cite{Zamolodchikov}. It is, however, challenging to go beyond the specific case of $K=1$ with this approach. We have found a way to achieve that by treating $K-1$ as a small expansion parameter. This enables us to perform a perturbative calculation near $K=1$, and obtain a set of asymptotically exact mean field equations for the two competing order parameters there. We are able to find the conditions under which a nontrivial solution with both orders coexisting near this self-dual point. Such a solution, if stable, would correspond to the coexistence of the DW and SF orders or a supersolid phase. However, energy comparison between the coexistence solution and the solutions of having a single nonzero order parameter shows that the former is energetically unfavorable. This suggests that the DW and SF phases are connected by a first-order transition when the system is very close to the self-dual point, and there is no supersolid phase.

The rest of the paper is organized as follows. We begin by reviewing the coupled-wire system and its Hamiltonian in Sec.~\ref{sec:app-K-1}. The interaction between neighboring wires is treated by the mean field approximation. When each wire has $K\approx 1$, we demonstrate explicitly how the mean field Hamiltonian can be transformed into the integrable sine-Gordon model. This allows us to derive a set of asymptotically exact mean field equations for the possible supersolid near the self-dual point in Sec.~\ref{sec:MF-sol}. Then, we obtain the corresponding solution, and also the solutions for pure DW and SF orders in the same section. Importantly, we identify the (narrow yet finite) region in which the supersolid solution exists. In order to determine the phase diagram of the system, we evaluate and compare the energies for different possible orders in Sec.~\ref{sec:en-analysis}. When the system is near the self-dual point, our numerical results show that the DW and SF orders always have lower energy than the supersolid order. The corresponding phase diagrams of the system at different effective interwire coupling strengths are also reported. Finally, we summarize our work and discuss its possible implications in Sec.~\ref{sec:conclusion}. Some technical details of calculation are given in the three Appendices.

\section{Approximation near $K=1$}
\label{sec:app-K-1}

The system that we consider is an array of one-dimensional wires (the words ``wires" and ``chains" will be used interchangeably), with the following Hamiltonian density~\cite{Tsvelik2002, Fradkin2010},
\begin{align} \label{eq:start}
\nonumber
\mathcal{H}
=&~\frac{v}{2}\sum_{i}
\left[K\left(\partial_x\theta_i\right)^2
+\frac{1}{K}\left(\partial_x\phi_i\right)^2\right]
\\ \nonumber
&-\sum_{\langle i,j\rangle} \mathcal{J}_S \cos{[\sqrt{2\pi}(\theta_i-\theta_j)]}
\\
&-\sum_{\langle i,j\rangle} 
\mathcal{J}_C \cos{[\sqrt{2\pi}(\phi_i-\phi_j)]}.
\end{align}
The first line of $\mathcal{H}$ describes an array of identical one-dimensional wires. Each of them is described by the Luttinger liquid model with the Luttinger parameter $K$. Its value is determined by the microscopic properties of the wire. The set of dual fields $\theta_i$ and $\phi_j$ satisfy the commutation relation, 
\begin{eqnarray}
[\phi_i(x), \partial_{x'}\theta_j(x')]=i\delta_{ij}\delta(x-x').
\end{eqnarray}
Physically, $\theta_i$ and $\phi_i$ stand for the phase fields of the superfluid and density wave along each wire. Nearest neighboring wires $i$ and $j$ are coupled by the Josephson and DW terms, with the respective coupling strengths $\mathcal{J}_S$ and $\mathcal{J}_C$. When $\mathcal{J}_S>0$, it favors a pinning for the term, $\cos{[\sqrt{2\pi}(\theta_i-\theta_j)}]=1$. Similarly, a pinning of $\cos{[\sqrt{2\pi}(\phi_i-\phi_j)}]=1$ is favored if $\mathcal{J}_C>0$. These two terms cannot be pinned simultaneously as $\phi_i$ and $\theta_i$ do not commute. This corresponds to a competition between the SF and DW orders. Notice that both $\mathcal{J}_S$ and $\mathcal{J}_C$ are bare coupling strengths. An ultraviolet energy cutoff scale $\Lambda$ will be introduced later when the vertex operators are normal ordered~\cite{Konik}. Notice that $\mathcal{H}$ is invariant under the transformation $\theta_i\leftrightarrow\phi_i$ when $K=1$ and 
$\mathcal{J}_S=\mathcal{J}_C$. This special point is known as the self-dual point in the literature~\cite{Nersesyan, Tsvelik2002, Fradkin2010}.

To understand the competition between the DW and SF orders, it is useful to discuss the scaling dimensions of the density wave and Josephson coupling terms in $\mathcal{H}$. From the two-point correlation functions:
\begin{align}
\langle\theta(\bm{x}_1)\theta(\bm{x}_2)\rangle
&\sim -\frac{1}{4\pi K}\ln{|\bm{x}_2-\bm{x}_1|^2},
\\
\langle\phi(\bm{x}_1)\phi(\bm{x}_2)\rangle
&\sim -\frac{K}{4\pi}\ln{|\bm{x}_2-\bm{x}_1|^2},
\end{align}
it is straightforward to deduce that
\begin{align}
\langle e^{i\sqrt{2\pi}\theta(\bm{x}_1)} e^{-i\sqrt{2\pi}\theta(\bm{x}_2)} \rangle
&\sim
\frac{1}{|\bm{x}_1-\bm{x}_2|^{1/K}},
\\
\langle e^{i\sqrt{2\pi}\phi(\bm{x}_1)} e^{-i\sqrt{2\pi}\phi(\bm{x}_2)} \rangle
&\sim
\frac{1}{|\bm{x}_1-\bm{x}_2|^K}.
\end{align}
Therefore, the two cosine terms (note that each term is a sum of products between \textit{two} vertex operators as $\theta_i$ and $\theta_j$ are decoupled under the mean field approximation) in the interchain coupling terms have scaling dimensions,
\begin{align} 
\label{eq:scaling-S}
\Delta_S
&=\frac{1}{2K}\times 2
=1/K,
\\
\label{eq:scaling-C}
\Delta_C
&=\frac{K}{2}\times 2
=K.
\end{align}
Therefore the RG equations for $\mathcal{J}_S$ and $\mathcal{J}_C$ are
\begin{align}
\frac{d\mathcal{J}_S}{d\ell}
&=(D-\Delta_S)\mathcal{J}_S,
\\
\frac{d\mathcal{J}_C}{d\ell}
&=(D-\Delta_C)\mathcal{J}_C,
\end{align}
where $D=d+1$ is the spacetime dimension. When $1/2<K<2$~\cite{footnote1}, $\Delta_S<2$ and $\Delta_C<2$. Hence, both interchain coupling terms are relevant in $1+1$-dimension. The case with $K=1$ is special because both terms have the same scaling dimension $\Delta=1$. Thus, a maximal competition between the two orders is realized there.

Applying the mean field approximation to the interwire coupling terms~\cite{Tsvelik2002}, the Hamiltonian density reduces to the form that describes a collection of decoupled single chains,
\begin{align} \label{eq:H_MF}
\nonumber
\mathcal{H}_{MF}
=&\sum_i
\left\{
\frac{v}{2}
\left[K\left(\partial_x\theta_i\right)^2
+\frac{1}{K}\left(\partial_x\phi_i\right)^2\right]\right.
\\
&\left. -g_S \cos{(\sqrt{2\pi}\theta_i)}
-g_C \cos{(\sqrt{2\pi}\phi_i)}
\right\}.
\end{align}
Here, $g_S$ and $g_C$ are the mean field parameters for the Josephson and density wave couplings, respectively. In principle, the above mean field approximation becomes exact in the limit of infinite co-dimension or infinite coordination number. We denote the coordination number as $z$. Then, $g_S$ and $g_C$ satisfy the self-consistency equations,
\begin{align}
\label{eq:self-gS}
g_S
&=z\mathcal{J}_{S}\langle \cos{(\sqrt{2\pi}\theta)\rangle},
\\
\label{eq:self-gC}
g_C
&=z\mathcal{J}_{C}\langle \cos{(\sqrt{2\pi}\phi)\rangle}.
\end{align}
For each decoupled wire, its effective Hamiltonian $h_i$ resembles the double sine-Gordon model that cannot be solved analytically in general. For the system to have a finite energy, both vertex operators in the field theory need to be normal ordered. The UV energy cutoff $\Lambda=1/a$ enters when the expectation values of the vertex operators are considered. As a result, the bare coupling constant is renormalized as~\cite{Konik}
\begin{eqnarray}
g\rightarrow \tilde{g}=g a^{\Delta_\beta}, 
\end{eqnarray}
where $\Delta_\beta$ is the scaling dimension of the vertex operator $:\cos{(\beta\theta)}:$. This leads to the dimension of the renormalized effective coupling constant, $[\tilde{g}]=E^{2-\Delta_\beta}$ where $E$ stands for energy.

\subsection{Rotation of order parameters at $K=1$}
\label{sec:rotation-order}

At $K=1$, both normal ordered vertex operators in Eq.~\eqref{eq:H_MF} have the same scaling dimension $\Delta=1/2$. Hence, the mean field Hamiltonian density for a single chain with the normal ordered vertex operators is
\begin{align} \label{eq:H_normal}
\nonumber
h_0
=
&\frac{v}{2}
\left[\left(\partial_x\theta\right)^2
+\left(\partial_x\phi\right)^2\right]
\\
&-g_S a^{1/2}:\cos{(\sqrt{2\pi}\theta)}:
-g_C a^{1/2} :\cos{(\sqrt{2\pi}\phi)}:.
\end{align}
In the following discussion, we will skip the normal ordering notation. All vertex operators below are normal ordered, unless otherwise specified. As 
already advertised in Sec.~\ref{sec:intro}, one can combine the two cosine terms into a single one by performing a rotation in the order parameter space,
\begin{eqnarray} \label{eq:transform}
\begin{pmatrix}
\cos{\sqrt{2\pi}\tilde{\theta}} \\ \cos{\sqrt{2\pi}\tilde{\phi}}   
\end{pmatrix}
=
\begin{pmatrix}
\cos{\alpha} & -\sin{\alpha} \\
\sin{\alpha} & \cos{\alpha}
\end{pmatrix}
\begin{pmatrix}
\cos{\sqrt{2\pi}\theta} \\ \cos{\sqrt{2\pi}\phi}
\end{pmatrix}.
\end{eqnarray}
By setting 
\begin{eqnarray} \label{eq:tan-alpha}
\tan{\alpha}=\frac{g_S}{g_C},
\end{eqnarray}
the two vertex operators are combined into
\begin{align}
\nonumber
&g_S a^{1/2}:\cos{(\sqrt{2\pi}\theta)}:
+g_C a^{1/2} :\cos{(\sqrt{2\pi}\phi)}:
\\
=&\sqrt{g_S^2+g_C^2}~a^{1/2} :\cos{(\sqrt{2\pi}\tilde{\phi})}:.
\end{align}
In order to determine $h_0$ in terms of $\tilde{\theta}$ and $\tilde{\phi}$, it is necessary to deduce how the derivative terms (i.e., the kinetic energy parts) transform.

Since the Luttinger liquid model is a conformal field theory with central charge $c=1$, the short distance behaviors of different conformal fields in the theory are dictated by the operator product expansion (OPE)~\cite{CFT-book}. This behavior should be independent of the transformation in Eq.~\eqref{eq:transform}. In terms of the original fields $\theta$ and $\phi$, we have the following OPE:
\begin{align} \label{eq:old-OPE}
\nonumber
&\lim_{x_2\rightarrow x_1}
\cos{[\sqrt{2\pi}{\theta}(x_1)]}\cos{[\sqrt{2\pi}{\theta}(x_2)]}
\\ \nonumber
=~&\frac{1}{2}\lim_{x_{12}\rightarrow 0}
\left\{
\frac{1}{x_{12}}
+x_{12}\left[\cos{[\sqrt{8\pi}\theta(x)]}-\pi[\partial_x\theta(x)]^2\right]\right\}
\\
&+\mathcal{O}(x_{12}^2).
\end{align}
In the above equation, we have defined $x_{12}=|x_2-x_1|$ and $x=x_1=x_2$. We can also evaluate the same OPE in terms of the new set of fields, $\tilde{\theta}$ and $\tilde{\phi}$. However, this cannot be done before determining $h_0$ after the transformation. This requires us to relate the current model with its corresponding microscopic model for spins~\cite{Nersesyan}. 

In fact, $h_0$ can be viewed as the continuum field theory limit of the spin-$1/2$ XXZ model under  staggered magnetic fields in both $x$ and $z$ directions. The spin system takes the Hamiltonian,
\begin{eqnarray}
H
=H_0-h_x\sum_j (-1)^j S_j^x-h_z\sum_j (-1)^j S_j^z,
\end{eqnarray}
where $H_0$ is the Hamiltonian of the XXZ model
\begin{eqnarray}
H_0
=J \sum_j \left[\left(S_j^x S_{j+1}^x + S_j^y S_{j+1}^y\right)
+\Delta S_j^z S_{j+1}^z\right].
\end{eqnarray}
With the convention that both bosonic fields $\phi(x)$ and $\theta(x)$ to have the same radius of compatification $R=1/\sqrt{2\pi}$, the spin operators can be bosonized as~\cite{Cabra, Grynberg}
\begin{align}
\label{eq:Fradkin-Sz}
S_z(x)
&=\frac{1}{\sqrt{2\pi}}\partial_x\phi(x)
+a_1\cos{[\sqrt{2\pi}\phi(x)+2k_F x]},
\\ \nonumber 
S_{\pm}(x)
&=(-1)^{x/a}
e^{\pm i\sqrt{2\pi}\theta(x)}
\\ \label{eq:Fradkin-Spm}
&\quad\quad\quad\times
\left\{b_1+b_2\cos{[\sqrt{2\pi}\phi(x)+2k_Fx]}\right\}.
\end{align}
After bosonization, $H$ takes the form of $h_0$ with $K$ being determined by the anisotropy term $\Delta$ in the XXZ model~\cite{footnote-Ukmlapp}. In particular, the Heisenberg point with $\Delta=1$ corresponds to $K=1$ in the Luttinger liquid model. The transformation in Eq.~\eqref{eq:transform} corresponds to a change in the spin axis on the $x-z$ plane for $H$. The term $\bm{S}_j\cdot\bm{S}_{j+1}$ in the Heisenberg model is invariant under the rotation of the spin axis. Hence, it is also expected that the derivative terms in $h_0$ is also invariant under the transformation in Eq.~\eqref{eq:transform}. As a result, we have in terms of the new set of fields $\tilde{\theta}$ and $\tilde{\phi}$,
\begin{align} \label{eq:H_new}
\nonumber
h_0
=&\frac{v}{2}
\left[(\partial_x\tilde{\theta})^2+(\partial_x\tilde{\phi})^2\right]
\\
&-\sqrt{g_S^2+g_C^2}~a^{1/2} :\cos{(\sqrt{2\pi}\tilde{\phi})}:.
\end{align}
A more detailed discussion can be found in Appendix~\ref{app:bose-XXZ}.

Now, we can recalculate the same OPE in Eq.~\eqref{eq:old-OPE} but in terms of $\tilde{\theta}$ and $\tilde{\phi}$. This gives
\begin{widetext}
\begin{align} \label{eq:after-OPE}
\nonumber
&\lim_{x_2\rightarrow x_1}
\left[\cos{\alpha}\cos{\sqrt{2\pi}\tilde{\theta}(x_1)}+\sin{\alpha}\cos{\sqrt{2\pi}\tilde{\phi}(x_1)}\right]
\left[\cos{\alpha}\cos{\sqrt{2\pi}\tilde{\theta}(x_2)}+\sin{\alpha}\cos{\sqrt{2\pi}\tilde{\phi}(x_2)}\right]
\\  \nonumber
=~&\left(\frac{1}{2}\cos^2{\alpha}\right)
\lim_{x_{12}\rightarrow 0}
\left\{
\frac{1}{x_{12}}
+x_{12}\left[\cos{[\sqrt{8\pi}\tilde{\theta}(x)]}-\pi[\partial_x\tilde{\theta}(x)]^2\right]+\mathcal{O}(x_{12}^2)
\right\}
\\
&+\left(\frac{1}{2}\sin^2{\alpha}\right)
\lim_{x_{12}\rightarrow 0}
\left\{
\frac{1}{x_{12}}
+x_{12}\left[\cos{[\sqrt{8\pi}\tilde{\phi}(x)]}-\pi[\partial_x\tilde{\phi}(x)]^2\right]+\mathcal{O}(x_{12}^2)
\right\}.
\end{align}
\end{widetext}
By comparing the terms with scaling dependence $x_{12}$ in Eqs.~\eqref{eq:old-OPE} and~\eqref{eq:after-OPE}, we find
\begin{align} 
\label{eq:theta-transform}
\left[\partial_x\theta(x)\right]^2
&=\cos^2{\alpha}~[\partial_x\tilde{\theta}(x) ]^2
+\sin^2{\alpha}~[\partial_x\tilde{\phi}(x)]^2,
\\ \label{eq:phi-transform}
\left[\partial_x\phi(x)\right]^2
&=\sin^2{\alpha}~[\partial_x\tilde{\theta}(x)]^2
+\cos^2{\alpha}~[\partial_x\tilde{\phi}(x)]^2.
\end{align}
Note that the above results preserve the SU(2) symmetry of $h_0$ at $K=1$ (c.f. the corresponding O(3) symmetry in the XXZ model at the Heisenberg point $\Delta=1$). 

\subsection{Perturbation near $K=1$}

Let us recall that the Hamiltonian density after the mean field approximation takes the form of Eq.~\eqref{eq:H_MF}. In the following discussion, we will only focus on the Hamiltonian density for a single chain, given by
\begin{align} 
\nonumber
h
=&
\frac{v}{2}
\left[K\left(\partial_x\theta\right)^2
+\frac{1}{K}\left(\partial_x\phi\right)^2\right]
\\
&-g_S \cos{(\sqrt{2\pi}\theta)}
-g_C \cos{(\sqrt{2\pi}\phi)}.
\end{align}
To go beyond the special case with $K=1$, we start by separating $h=h_0+\delta h$. Here, $h_0$ is given by Eq.~\eqref{eq:H_normal} which is the Hamiltonian density for a single decoupled chain when $K=1$. After the transformation in Eq.~\eqref{eq:transform}, Sec.~\ref{sec:rotation-order} shows that $h_0$ would take the form of Eq.~\eqref{eq:H_new}. 

Now, we consider the deviation term $\delta h$. Since we only focus on the system that is very close to the self-dual point, we have $K=1+\epsilon$ and $\epsilon\ll 1$. This enables us to expand $\delta h$ in the first order of $\epsilon$, which gives
\begin{align} \label{eq:perturbation}
\nonumber
\delta h
&=\frac{v}{2}\left[(K-1)(\partial_x\theta)^2+\left(\frac{1}{K}-1\right)(\partial_x\phi)^2\right]
\\
&=\frac{\epsilon v}{2}\left[(\partial_x\theta)^2-(\partial_x\phi)^2\right]+\mathcal{O}(\epsilon^2).
\end{align}
Strictly speaking, the rotation of order parameters in Eq.~\eqref{eq:transform} can only be performed at $K=1$. In Appendix~\ref{app:bose-XXZ}, we carry out the corresponding transformation in the XXZ model. The challenge in deducing the corresponding Luttinger model from bosonizing the transformed XXZ model is also discussed there. Here, we argue that Eqs.~\eqref{eq:theta-transform} and~\eqref{eq:phi-transform}  remain asymptotically exact near $K=1$. In other words, they still provide a good approximation to the transformations of the conformal fields when $\epsilon\ll 1$. Using those equations, we express $\delta h$ in terms of the transformed fields $\tilde{\theta}$ and $\tilde{\phi}$:
\begin{eqnarray}
\delta h
=\frac{\epsilon v}{2}
\left(\cos^2\alpha-\sin^2\alpha\right)
\left[(\partial_x\tilde{\theta})^2-(\partial_x\tilde{\phi})^2\right].
\end{eqnarray}
Let us recall that $\tan{\alpha}=g_S/g_C$ [see Eq.~\eqref{eq:tan-alpha}], in order to combine the two cosine terms. Therefore, $\sin^2\alpha=g_S^2/(g_S^2+g_C^2)$ and 
$\cos^2\alpha=g_C^2/(g_S^2+g_C^2)$. Using this, we finally obtain
\begin{align} \label{eq:delta-h}
\nonumber
\delta h
&=\frac{\epsilon v}{2}\left(\frac{g_C^2-g_S^2}{g_C^2+g_S^2}\right)
\left[(\partial_x\tilde{\theta})^2-(\partial_x\tilde{\phi})^2\right]
\\
&=\frac{\epsilon v}{2}G\left[(\partial_x\tilde{\theta})^2-(\partial_x\tilde{\phi})^2\right].
\end{align}
For later convenience, we have defined the dimensionless parameter,
\begin{eqnarray}
G=\frac{g_C^2-g_S^2}{g_C^2+g_S^2}.
\end{eqnarray}
It must satisfy $-1\leq G\leq 1$. For $g_S=0$ and $g_C=0$, they correspond respectively to $G=1$ and $G=-1$. 

By combining Eqs.~\eqref{eq:H_new} and~\eqref{eq:delta-h}, we claim that the system near $K=1$ takes the approximate Hamiltonian density,
\begin{align}
\nonumber
h_0+\delta h
=~&\frac{v}{2}\left[\left(1+\epsilon G\right)(\partial_x\tilde{\theta})^2
+\left(1-\epsilon G\right)(\partial_x\tilde{\phi})^2\right]
\\
&-\sqrt{g_S^2+g_C^2}~ a^{\Delta'}
:\cos{(\sqrt{2\pi}\tilde{\phi})}:.
\end{align}
Note that $\Delta' \neq 1/2$. It is because the scaling dimension of the cosine term changes after the introduction of the perturbation $\delta h$. To rewrite $h_0+\delta h$ in the form of a standard sine-Gordon model, we introduce a set of rescaled fields,
\begin{align}
\vartheta
&=\sqrt{1+\epsilon G}~\tilde{\theta},
\\
\varphi
&=\sqrt{1-\epsilon G}~\tilde{\phi}.
\end{align}
As $-1\leq G \leq 1$ and $|\epsilon|\ll 1$, one has $|\epsilon G|\ll 1$. The commutator 
$[\tilde{\phi}(x), \partial_{x}\tilde{\theta}(x')]=i\delta(x-x')$ leads to
$[\varphi(x), \partial_x\vartheta(x')]=i\delta(x-x')+\mathcal{O}[(\epsilon G)^2]$. Thus, the canonical  quantization rule holds in the first order of $\epsilon$, and remains a good approximation near $K=1$. In terms of the rescaled fields, 
\begin{align} \label{eq:eff-SG-model}
\nonumber
h_0+\delta h
=&\frac{v}{2}
\left[(\partial_x\vartheta)^2+(\partial_x\varphi)^2\right]
\\
&-\sqrt{g_S^2+g_C^2}~ a^{\Delta'}
\cos{\left(\sqrt{\frac{2\pi}{1-\epsilon G}}~\varphi\right)}.
\end{align}
In this form, it is straightforward to deduce that
\begin{eqnarray} \label{eq:scaling-dim}
\Delta'
=\frac{1}{4\pi}
\left(\sqrt{\frac{2\pi}{1-\epsilon G}}\right)^2
=\frac{1}{2(1-\epsilon G)}.
\end{eqnarray}

\section{Solutions of mean field equations}
\label{sec:MF-sol}

Since the sine-Gordon model is an integrable model, the form of $h_0+\delta h$ in Eq.~\eqref{eq:eff-SG-model} allows us to study the competition and coexistence of different quantum phases in the coupled wire system near the self-dual point quantitatively. Consider the standard sine-Gordon model with the action~\cite{Coleman1975},
\begin{eqnarray} \label{eq:SGmodel}
S=\int d^2x \left[\frac{1}{2}(\partial_\mu\phi)^2+\mu_{SG}:\cos{(\alpha\phi)}:\right].
\end{eqnarray}
It is noted that the renormalized coupling constant $\mu_{SG}$ has the dimension $E^{2-\Delta_\alpha}$, where $\Delta_\alpha$ is the scaling dimension of the normal ordered vertex operator. In the present case, the latter depends on $\epsilon G$ [see Eq.~\eqref{eq:scaling-dim}]. The spectrum of the sine-Godron model is controlled by the soliton mass $M$, which has been determined analytically~\cite{Zamolodchikov, Konik}:
\begin{eqnarray} \label{eq:M-SG}
M
=\frac{2\Gamma\left(\frac{\xi_{SG}}{2}\right)}{\sqrt{\pi}\Gamma\left(\frac{1+\xi_{SG}}{2}\right)}
\left[\frac{\mu_{SG}\pi\Gamma\left(1-\frac{\alpha^2}{8\pi}\right)}{2\Gamma\left(\frac{\alpha^2}{8\pi}\right)}\right]^{\frac{1}{2-\frac{\alpha^2}{4\pi}}}.
\end{eqnarray}
Here, the symbol $\Gamma(x)$ denotes the Gamma function. The parameter $\xi_{SG}$ is defined as
\begin{eqnarray}
\xi_{SG}
=\frac{\alpha^2}{8\pi-\alpha^2}.
\end{eqnarray}
The formula for $M$ is physical only when $0\leq \alpha^2\leq 8\pi$~\cite{Konik}. The associated ground state energy density of the sine-Gordon model is related to its soliton mass by
\begin{eqnarray} \label{eq:gs-en}
\mathcal{E}
=-\frac{M^2}{4}\tan{\left(\frac{\pi\xi_{SG}}{2}\right)}.
\end{eqnarray}
This energy density remains finite when $0\leq \xi_{SG}< 1$ (i.e., $0\leq\alpha^2< 4\pi$).

For our present case, the corresponding parameters take values
\begin{align}
\alpha
&=\sqrt{\frac{2\pi}{1-\epsilon G}},
\\
\mu_{SG}
&=\sqrt{(g_S^2+g_C^2)a^{1/(1-\epsilon G)}},
\\
\xi_{SG}
&=\frac{1}{3-4\epsilon G}.
\end{align}
Since $|\epsilon G|\ll 1$, $\alpha^2\approx 2\pi$. Hence, both Eqs.~\eqref{eq:M-SG} and~\eqref{eq:gs-en} are well defined in the present discussion. A direct substitution leads to the ground state energy density for general values of $g_S$ and $g_C$ as
\begin{widetext}
\begin{align} \label{eq:E_Z}
\nonumber
\mathcal{E}(Z)
&=-\Lambda^2
\left[\tan{\left(\frac{\pi}{6-8Z}\right)}\right]
\left(\frac{\pi}{16^{1-Z}}\right)^{\frac{1}{3-4Z}}
\left[\frac{\Gamma\left(\frac{1}{6-8Z}\right)}
{\Gamma\left(\frac{2-2Z}{3-4Z}\right)}\right]^2
\left[
\frac{\Gamma\left(\frac{3-4Z}{4-4Z}\right)}
{\Gamma\left(\frac{1}{4-4Z}\right)}
\right]^{\frac{4-4Z}{3-4Z}}
\left(\frac{g_S^2+g_C^2}{\Lambda^4}\right)^{\frac{2(1-Z)}{3-4Z}}
\\
&=-\Lambda^2 F(Z)\left(\frac{g_S^2+g_C^2}{\Lambda^4}\right)^{\frac{2(1-Z)}{3-4Z}}.
\end{align}
\end{widetext}
Here, the dimensionless parameter $Z=\epsilon G$. Also notice that both $g_S/\Lambda^2$ and $g_C/\Lambda^2$ are dimensionless quantities. Thus, $\mathcal{E}(Z)$ has the correct dimension $E^2$, which is consistent with being the energy density of a single effective one-dimensional wire. 

\subsection{Solutions for pure DW and SF orders}

In principle, the analytical form of $\mathcal{E}(Z)$ allows one to formulate a set of self-consistency equations to describe the competition and possible coexistence of DW and SF orders in the coupled wire system 
at $|\epsilon|\ll 1$. This will be discussed in Sec.~\ref{sec:coexistence}. Here, we first revisit the two simpler cases in which either one of $g_S$ or $g_C$ is zero~\cite{Tsvelik2002}. They describe the scenarios that the system has a pure density wave or superfluid order, respectively. In both cases, it is not necessary to perform any rotation in the mean field Hamiltonian as it already has the form of standard sine-Gordon model.

First, we consider the case with $g_S=0$ but $g_C\neq 0$. In this case, the self-consistency equation in Eq.~\eqref{eq:self-gS} is satisfied automatically by having both sides equal to zero. The mean field Hamiltonian density reduces to the standard sine-Gordon model. By defining $\theta'=\sqrt{K}\theta$ and $\phi'=\phi/\sqrt{K}$, one has
\begin{align}
\nonumber
h_{\text{DW}}
=&\frac{v}{2}
\left[\left(\partial_x\theta'\right)^2
+\left(\partial_x\phi'\right)^2\right]
\\
&-\frac{g_C}{\Lambda^{\Delta_C}} :\cos{\left(\sqrt{2\pi K}\phi'\right)}:,
\end{align}
where $\Delta_C=K/2$ is the scaling dimension of the vertex operator at a generic value of $K$. Following the above discussion of sine-Gordon model, the corresponding mean field energy density is
\begin{eqnarray}
\mathcal{E}_{\text{DW}}
=-\Lambda^2 F\left(\frac{K-1}{K}\right)
\left(\frac{g_C^2}{\Lambda^4}\right)^{2/(4-K)}.
\end{eqnarray}
Applying the Hellmann-Feynman theorem, one obtains
\begin{align}
\nonumber
\langle \cos{(\sqrt{2\pi}\phi)\rangle}
&=-\frac{\partial \mathcal{E}_{\text{DW}}}{\partial g_C}
\\
&=\left(\frac{4}{4-K}\right) 
F\left(\frac{K-1}{K}\right)
\left(\frac{g_C}{\Lambda^2}\right)^{\frac{K}{4-K}}.
\end{align}
Thus, the corresponding solution to Eq.~\eqref{eq:self-gC} is given by
\begin{eqnarray} \label{eq:gC-exact}
\frac{g_C^2}{\Lambda^4}
=\left[
\left(\frac{4z}{4-K}\right) F\left(\frac{K-1}{K}\right) \left(\frac{\mathcal{J}_C}{\Lambda^2}\right)
\right]^{\frac{4-K}{2-K}}.
\end{eqnarray}

Since $Z=\epsilon (g_C^2-g_S^2)/(g_C^2+g_S^2)=\epsilon$ when $g_S=0$, one may wonder whether is it possible to substitute $Z=\epsilon$ in Eq.~\eqref{eq:E_Z} and obtain the corresponding mean field solution of the density wave order. This will lead to the solution
\begin{eqnarray}
\frac{g_C^2}{\Lambda^4}
\approx
\left[
4z\left(\frac{1-\epsilon}{3-4\epsilon}\right)F(\epsilon)
\left(\frac{\mathcal{J}_C}{\Lambda^2}\right)
\right]^{\frac{3-4\epsilon}{1-2\epsilon}}.
\end{eqnarray}
It is straightforward to check that the result is consistent with Eq.~\eqref{eq:gC-exact} up to the first order of $\epsilon$. This is because the perturbation term $\delta h$ in Eq.~\eqref{eq:perturbation} has already been truncated in the first order in $\epsilon$, so does the energy density derived from it. In the later part of the work, we will only focus on the solution in Eq.~\eqref{eq:gC-exact}.

When $g_C=0$, the mean field Hamiltonian density also reduces to the standard sine-Gordon model,
\begin{align}
h_{\text{SF}}
=\frac{v}{2}
\left[\left(\partial_x\theta'\right)^2
+\left(\partial_x\phi'\right)^2\right]
-\frac{g_S}{\Lambda^{\Delta_S}} :\cos{\left(\sqrt{\frac{2\pi}{K}}\theta'\right)}:.
\end{align}
Here, $\Delta_S=1/(2K)$. The corresponding mean field energy density is
\begin{eqnarray}
\mathcal{E}_{\text{SF}}
=-\Lambda^2 F\left(1-K\right)
\left(\frac{g_S^2}{\Lambda^4}\right)^{2K/(4K-1)}.
\end{eqnarray}
It leads to the solution of Eq.~\eqref{eq:self-gS},
\begin{eqnarray} \label{eq:gS-exact}
\frac{g_S^2}{\Lambda^4}
=\left[
\left(\frac{4zK}{4K-1}\right) F\left(1-K\right) \left(\frac{\mathcal{J}_S}{\Lambda^2}\right)
\right]^{\frac{4K-1}{2K-1}}.
\end{eqnarray}
The solutions in Eq.~\eqref{eq:gC-exact} and~\eqref{eq:gS-exact} obey the duality relation: $K\rightarrow 1/K$, $\mathcal{J}_S\leftrightarrow\mathcal{J}_C$, and $g_S\leftrightarrow g_C$~\cite{Tsvelik2002}.

\subsection{Solutions for coexisting DW and SF orders}
\label{sec:coexistence}

An important feature of intertwined orders is that they may actually coexist instead of 
suppressing each other. To study this possibility, one can in principle formulate a set of self-consistency equations by directly applying the Hellmann-Feynman theorem on $\mathcal{E}(Z)$ in Eq.~\eqref{eq:E_Z} to obtain $\langle \cos{(\sqrt{2\pi}\theta)\rangle}$ and $\langle \cos{(\sqrt{2\pi}\phi)\rangle}$ in Eqs.~\eqref{eq:self-gS} and~\eqref{eq:self-gC}. However, this is a very challenging task as $\mathcal{E}(Z)$ is a complicated function of $g_S$ and $g_C$ when none of them is zero. Instead, we first make an approximation for $\mathcal{E}(Z)$ by expanding it up to the first order in $\epsilon$. This leads to
\begin{align} \label{eq:E-1}
\nonumber
\mathcal{E}(\epsilon)
=~&\mathcal{E}_0
\times \left\{1+\frac{2\epsilon}{9}\left(\frac{g_C^2-g_S^2}{g_C^2+g_S^2}\right)
\ln{\left[
\varsigma
\left(\frac{g_S^2+g_C^2}{\Lambda^4}\right)\right]}
\right\}
\\
&+\mathcal{O}[(\epsilon G)^2].
\end{align}
Here, $\mathcal{E}_0$ is the corresponding energy density at $\epsilon=0$,
\begin{eqnarray} \label{eq:E-0}
\mathcal{E}_0
=-\eta\Lambda^2\left(\frac{g_S^2+g_C^2}{\Lambda^4}\right)^{2/3}.
\end{eqnarray}
In Eqs.~\eqref{eq:E-1} and~\eqref{eq:E-0}, we have defined the constants,
\begin{align} 
\label{eq:varsigma-def}
\varsigma
&=\frac{8\pi^2 e^{3\gamma}[\Gamma(3/4)]^2}{[\Gamma(1/4)]^2},
\\
\label{eq:eta-def}
\eta
&=\frac{1}{\sqrt{3}}\left(\frac{\pi}{16}\right)^{1/3}
\left[\frac{\Gamma(1/6)}{\Gamma(2/3)}\right]^2
\left[\frac{\Gamma(3/4)}{\Gamma(1/4)}\right]^{4/3},
\end{align}
where $\gamma\approx 0.577$ is the Euler constant.

Now, we apply the Hellmann-Feynman theorem on Eq.~\eqref{eq:E-1} to derive a set of approximate self-consistency equations that are valid for $|\epsilon| \ll 1$:
\begin{widetext}
\begin{align} 
\label{eq:MF-S}
g_S
&=\frac{4z\eta j_S}{3}\left[\frac{g_S}{(\kappa_S^2+\kappa_C^2)^{1/3}}\right]
\left\{
1+\frac{\epsilon}{3}\left(\frac{\kappa_C^2-\kappa_S^2}{\kappa_C^2+\kappa_S^2}\right)
-\frac{2\epsilon}{9}\left(\frac{2\kappa_C^2+\kappa_S^2}{\kappa_C^2+\kappa_S^2}\right)
\ln{
\left[\varsigma\left(\kappa_S^2+\kappa_C^2\right)\right]}
\right\},
\\
\label{eq:MF-C}
g_C
&=
\frac{4z\eta j_C}{3}\left[\frac{g_C}{(\kappa_S^2+\kappa_C^2)^{1/3}}\right]
\left\{
1+\frac{\epsilon}{3}\left(\frac{\kappa_C^2-\kappa_S^2}{\kappa_C^2+\kappa_S^2}\right)
+\frac{2\epsilon}{9}\left(\frac{\kappa_C^2+2\kappa_S^2}{\kappa_C^2+\kappa_S^2}\right)
\ln{
\left[\varsigma\left(\kappa_S^2+\kappa_C^2\right)\right]}
\right\}.
\end{align}
\end{widetext}
The set of dimensionless parameters are defined,
\begin{eqnarray}
\kappa_S=\frac{g_S}{\Lambda^2}
~,~
\kappa_C=\frac{g_C}{\Lambda^2}
~,~
j_S=\frac{\mathcal{J}_S}{\Lambda^2}
~,~
j_C=\frac{\mathcal{J}_C}{\Lambda^2}.
\end{eqnarray}

In order for the superfluid and density wave orders to coexist, both $g_S$ and $g_C$ need to be nonzero. At $\epsilon=0$ (or $K=1$), this is possible only when $\mathcal{J}_S=\mathcal{J}_C=\mathcal{J}$. This agrees with the previous result from interchain mean field approximation~\cite{Tsvelik2002}. In this case, there is only one self-consistency equation for $g_S$ and $g_C$. Its solution is
\begin{eqnarray} \label{eq:sol-e0}
\frac{g_S^2+g_C^2}{\Lambda^4}
=\left(\frac{4z\eta\mathcal{J}}{3\Lambda^2}\right)^3.
\end{eqnarray}
The solution suggest that $g_S$ and $g_C$ cannot be determined uniquely. This agrees with the high symmetry of the self-dual point at which the order parameter can point in any direction, but the overall norm is fixed by $\mathcal{J}$. Note that $\mathcal{J}>0$ according to our definition in Eq.~\eqref{eq:start}.

Going away from the self-dual point, we now solve Eqs.~\eqref{eq:MF-S} and~\eqref{eq:MF-C} for $\epsilon\ne 0$. By introducing two new variables $R=\kappa_C^2+\kappa_S^2$ and $r=\kappa_C^2-\kappa_S^2$, we can rewrite Eqs.~\eqref{eq:MF-S} and~\eqref{eq:MF-C} in the following form,
\begin{align} 
\frac{4z\eta j_S}{3R^{1/3}}
\left[
1+\frac{\epsilon}{3}\left(\frac{r}{R}\right)
-\frac{\epsilon}{9}\left(\frac{3R+r}{R}\right)
\ln{(\varsigma R)}
\right]
&=1,
\\
\frac{4z\eta j_C}{3R^{1/3}}
\left[
1+\frac{\epsilon}{3}\left(\frac{r}{R}\right)
+\frac{\epsilon}{9}\left(\frac{3R-r}{R}\right)
\ln{(\varsigma R)}
\right]
&=1.
\end{align}
From the first equation, we obtain an equation for $r$ in terms of $R$:
\begin{eqnarray}
r=\frac{3R}{4z\eta j_S\epsilon}
\left[\frac{9R^{1/3}-12z\eta j_S + 4z\eta j_S \epsilon \ln{(\varsigma R)}}
{3-\ln{(\varsigma R)}}
\right].
\end{eqnarray}
Substituting the above relation into the second equation, we obtain an equation for $R$:
\begin{eqnarray} \label{eq:ln-negative}
(\varsigma R)^{-1/3}\ln{\left[(\varsigma R)^{-1/3}\right]}
=\frac{3}{8z\eta\varsigma^{1/3}\epsilon}\left(\frac{1}{j_S}-\frac{1}{j_C}\right).
\end{eqnarray}
Let $y=\ln{\left[(\varsigma R)^{-1/3}\right]}$. Then, the above equation can be rewritten in the form:
\begin{eqnarray} 
ye^y
=\frac{3}{8z\eta\varsigma^{1/3}\epsilon}\left(\frac{1}{j_S}-\frac{1}{j_C}\right).
\end{eqnarray}
This can be solved analytically,
\begin{eqnarray}
y
=W_0\left[\frac{3}{8z\eta\varsigma^{1/3}\epsilon}\left(\frac{1}{j_S}-\frac{1}{j_C}\right)\right]
=W_0(X).
\end{eqnarray}
Here, we have assumed $(j_C-j_S)/\epsilon>0$, so that the equation for $y$ can be completely solved by the principal branch of the Lambert $W$ function denoted as $W_0(x)$~\cite{footnote-productlog}. Notice that the left hand side of Eq.~\eqref{eq:ln-negative} involves $\ln(\varsigma R)$. Since $0<R=\kappa_S^2+\kappa_C^2=(g_S^2+g_C^2)/\Lambda^4 \ll 1$ in the field theory limit, so $\ln{(\varsigma R)}$ is believed to be negative. Thus, $\epsilon$ and $j_C-j_S$ should have the same sign. This justifies our assumption. Furthermore, the symbol $X$ is introduced for later convenience,
\begin{eqnarray} \label{eq:def-X}
X
=X(j_S, j_C, \epsilon)
=\frac{3}{8z\eta\varsigma^{1/3}\epsilon}\left(\frac{1}{j_S}-\frac{1}{j_C}\right).
\end{eqnarray}
From the solution of $y$, we can obtain the solution of 
$R$:
\begin{align} \label{eq:R}
\nonumber
R
&=\frac{1}{\varsigma} 
\exp{\left[-3W_0\left(\frac{3}{8z\eta\varsigma^{1/3}\epsilon}
\left(\frac{1}{j_S}-\frac{1}{j_C}\right)\right)\right]}
\\
&=\left(\frac{8z\eta j_S}{3}\right)^3
\left(\frac{\epsilon}{1-J}\right)^3
\left[W_0\left(X\right)\right]^3.
\end{align}
Note that we have used the property $e^{nW(x)}=[x/W(x)]^n$ to obtain the second equality. We have also defined the dimensionless ratio $J=\mathcal{J}_S/\mathcal{J}_C$. The value of $R$ at $\epsilon=0$ and $J=1$ cannot be determined from the expression.

Using the result of $R$, we can solve for $r$,
\begin{align}
r
=3j_S^3\left[\frac{8z\eta W_0(X)}{3(J-1)}\right]^3 
\left[\frac{1+\epsilon \frac{J+1}{J-1} W_0(X) }{1+W_0(X)}\right]\epsilon^2.
\end{align}
Finally, we can solve for $\kappa_S^2$ and $\kappa_C^2$:
\begin{widetext}
\begin{align}
\label{eq:sol-ks2}
\kappa_S^2
&=\frac{j_S^3}{2}\left[\frac{8z\eta  W_0(X)}{3}\right]^3 \left(\frac{1}{1-J}\right)^4
\left[\frac{(3+\epsilon)(1-J)-2\epsilon (1+2J) W_0(X)}{1+W_0(X)}\right]\epsilon^2,
\\
\label{eq:sol-kc2}
\kappa_C^2
&=\frac{j_S^3}{2}\left[\frac{8z\eta  W_0(X)}{3}\right]^3 \left(\frac{1}{1-J}\right)^4
\left[\frac{(-3+\epsilon)(1-J)+2\epsilon (2+J) W_0(X)}{1+W_0(X)}\right]\epsilon^2.
\end{align}
\end{widetext}
The above solution obeys the duality relation: $\epsilon\rightarrow -\epsilon$, $j_S\leftrightarrow j_C$, and $\kappa_S^2\leftrightarrow\kappa_C^2$. In the original Hamiltonian, the duality relation is $K\rightarrow 1/K$, $\mathcal{J}_S\leftrightarrow\mathcal{J}_C$, and $\theta\leftrightarrow\phi$. The coexistence solution is physical if and only if both $\kappa_S^2$ and $\kappa_C^2$ are positive. This requires
\begin{align} \label{eq:sol-region}
\frac{(1-J)(3-\epsilon)}{2(2+J)}
<\epsilon W_0\left[\frac{3(1-J)}{8z\eta\varsigma^{1/3}\epsilon j_S }\right]
<\frac{(1-J)(3+\epsilon)}{2(1+2J)}.
\end{align}
The region in which the inequality is satisfied can be determined numerically. Since our focus is the region with $\epsilon\approx 0$ and $J\approx 1$, it is possible to make further approximations to Eq.~\eqref{eq:sol-region}. This is done in Appendix~\ref{app:approx-region}. The end result is given by Eq.~\eqref{eq:region1}, which indicates that the supersolid solution can only exist in a narrow region around the self-dual point. Furthermore, it is necessary to compare the energies between different possible solutions, so that the phase diagram of the system near the self-dual point can be determined.

\section{Energy analysis of different possible solutions}
\label{sec:en-analysis}

\begin{figure} [htb]
\includegraphics[width=3.5in]{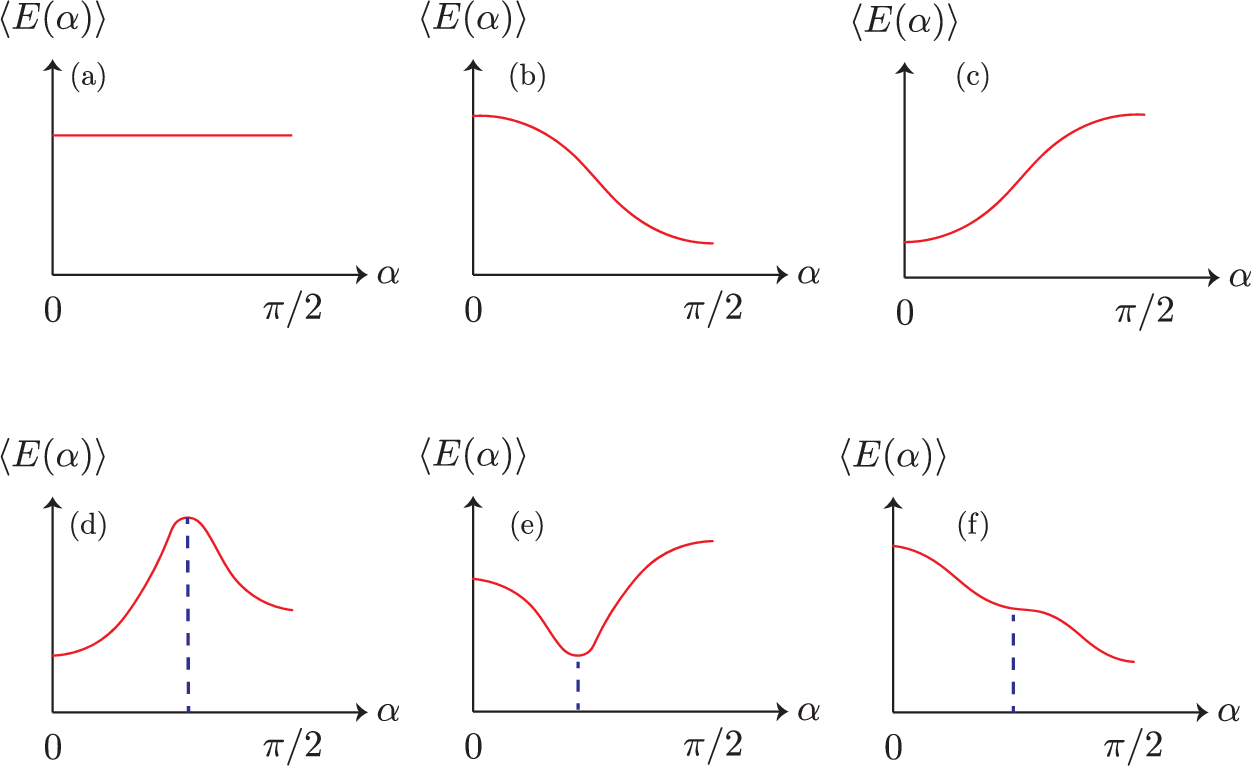}
\caption{Possible scenarios for the energy landscape of the system. Here, $\tan{\alpha}=\Upsilon_S/\Upsilon_C$ measures the ratio between the amplitudes of the superfluid and density wave order parameters, which are denoted as $\Upsilon_S$ and $\Upsilon_C$, respectively. The existence of a mean-field solution with coexisting order parameters (marked by the blue dashed lines) corresponds to having a local extremum or a highly unlikely point of inflection for $\langle E(\alpha)\rangle$ at an intermediate value of $\alpha_c\in(0, \pi/2)$.}
\label{fig:landscape}
\end{figure}

The self-dual point is so special that the order parameters can be rotated into each other while the energy of the system remains unchanged. There, only the overall amplitude $R=\kappa_S^2+\kappa_C^2$ is fixed by the coupling strength $\mathcal{J}=\mathcal{J}_S=\mathcal{J}_C$ as shown in Eq.~\eqref{eq:sol-e0}. Suppose we define an angle $\alpha$, such that $\tan{\alpha}=\Upsilon_S/\Upsilon_C$ measures the ratio between the amplitudes of the two order parameters. Then, the energy landscape of the system at the self-dual point with different values of $\alpha$ looks flat. This is illustrated schematically in Fig.~\ref{fig:landscape}(a).

Going away from the self-dual point, the mean field equations may or may not have a solution that describes the coexistence of the density wave and superfluid parameters. When such a coexistence is absent, our energy analysis below shows that the system will always develop an ordered state, either having a pure density wave order ($\alpha=0$) or a pure superfluid order ($\alpha=\pi/2$). Which one has a lower energy depends on the values of $K=1+\epsilon$, $\mathcal{J}_S$, and $\mathcal{J}_C$. Thus, the energy landscape of the system resembles either Fig.~\ref{fig:landscape}(b) or~\ref{fig:landscape}(c). Meanwhile, we did find solutions for coexisting orders in a certain region of the system parameters. This solution can correspond to a local minimum, local maximum, or a very unlikely point of inflection for the energy of the system. These three different scenarios are illustrated in Figs.~\ref{fig:landscape}(d)--(f).

To determine whether the coexistence of orders is energetically favorable, we evaluate the expectation value of the original Hamiltonian in Eq.~\eqref{eq:start},
\begin{align} 
\nonumber
\langle \mathcal{H}\rangle
=&~\frac{1}{2}\sum_{i}
\left[K\langle\left(\partial_x\theta_i\right)^2\rangle
+\frac{1}{K}\langle\left(\partial_x\phi_i\right)^2\rangle\right]
\\ \nonumber
&-\frac{z}{2}\sum_i
\left\{
\mathcal{J}_S \langle\cos{(\sqrt{2\pi}\theta_i)}\rangle\langle\cos{(\sqrt{2\pi}\theta)}\rangle\right.
\\
&\quad\quad\quad
+\left.
\mathcal{J}_C  \langle\cos{(\sqrt{2\pi}\phi_i)}\rangle\langle\cos{(\sqrt{2\pi}\phi)}\rangle
\right\},
\end{align}
under different solutions of the self-consistency equations. Here, we have set $v=1$ and made use of the fact that the mean field state is a product state of decoupled wires. This allows us to factorize the expectation value for the coupling terms $\cos{\sqrt{2\pi}(\theta_i-\theta_j)}$. In the following discussion, we only focus on a single effective chain $i$. For the coupling terms (or the potential energy term), the energy density is
\begin{eqnarray}
\langle U_i\rangle
=-\frac{z}{2}
\left[\mathcal{J}_S\left(-\frac{\partial\mathcal{E}}{\partial g_S}\right)^2
+\mathcal{J}_C\left(-\frac{\partial\mathcal{E}}{\partial g_C}\right)^2
\right].
\end{eqnarray}

For the expectation value of the derivative terms (or the kinetic energy term), we obtain it in the following way. We know that the Hamiltonian density of the system after the mean field approximation can be eventually written as a sine-Gordon model in Eq.~\eqref{eq:SGmodel}. The ground state energy density of the model is given by Eq.~\eqref{eq:gs-en}, which we denote it as $\langle h_{SG}\rangle$. Since 
$\langle h_{SG}\rangle=\langle K_i\rangle -\mu_{SG}\langle \cos(\alpha\varphi)\rangle$, it leads to
\begin{align}
\nonumber
\langle K_i\rangle
&=\langle h_{SG}\rangle + \mu_{SG}\langle \cos(\alpha\varphi)\rangle
\\ \nonumber
&=\mathcal{E}+ \mu_{SG}\left(-\frac{\partial\mathcal{E}}{\partial \mu_{SG}}\right)
\\
&=\left[1-\frac{2}{2-\alpha^2/(4\pi)}\right]\mathcal{E}.
\end{align}
It is reminded again that $\Delta=\alpha^2/(4\pi)$ is the scaling dimension of the vertex operator defined in Eq.~\eqref{eq:scaling-dim}. Therefore, one gets
\begin{align}
\nonumber
\frac{\langle \mathcal{H}_i\rangle}{\Lambda^2}
=&\left(\frac{\Delta}{\Delta-2}\right)
\frac{\mathcal{E}}{\Lambda^2}
\\
&-\frac{z}{2}
\left[
\frac{\mathcal{J}_S}{\Lambda^2}
\left(\frac{\partial\mathcal{E}}{\partial g_S}\right)^2
+\frac{\mathcal{J}_C}{\Lambda^2}
\left(\frac{\partial\mathcal{E}}{\partial g_C}\right)^2
\right].
\end{align}
In the following discussion, we will consider the dimensionless energy density $\langle \mathcal{H}_i\rangle/\Lambda^2$ instead of $\langle \mathcal{H}_i\rangle$. This enables us to ignore the UV cutoff dependence of the energy, and focus on its scaling behavior of the dimensionless interchain coupling strengths. From the above expression, one can verify that the solution in Eq.~\eqref{eq:sol-e0} minimizes
$\langle \mathcal{H}_i\rangle/\Lambda^2$ at $\epsilon=0$ when $\mathcal{J}_S=\mathcal{J}_C$. In addition, the corresponding expectation value can be evaluated,
\begin{eqnarray} \label{eq:en-K=1}
\nonumber
\left.\frac{\langle \mathcal{H}_i\rangle}{\Lambda^2}\right|_{\epsilon=0}
=-\left(\frac{\eta}{3}\right)^3\left(\frac{4z\mathcal{J}}{\Lambda^2}\right)^2,
\end{eqnarray}
where $\eta$ was defined in Eq.~\eqref{eq:eta-def}.

\subsection{Pure density wave or superfluid order} \label{sec:sol-enot0}

Now, we analyze $\langle \mathcal{H}_i\rangle/\Lambda^2$ when the mean field parameters $g_S$ and $g_C$ take values from the possible solutions that describe different orders in the system. From Eqs.~\eqref{eq:self-gS} and~\eqref{eq:self-gC}, one has
\begin{align} \label{eq:exp-h}
\left.\frac{\langle \mathcal{H}_i\rangle}{\Lambda^2}\right|_{\text{sol}}
=\left(\frac{\Delta}{\Delta-2}\right)\frac{\mathcal{E}}{\Lambda^2}
-\frac{1}{2z}
\left(\frac{g_S^2/\Lambda^4}{\mathcal{J}_S/\Lambda^2}
+\frac{g_C^2/\Lambda^4}{\mathcal{J}_C/\Lambda^2}
\right).
\end{align}
Note that the subscript ``sol" emphasizes that Eq.~\eqref{eq:exp-h} is valid only when $g_S$ and $g_C$ satisfy the self-consistency equations, Eqs.~\eqref{eq:self-gS} and~\eqref{eq:self-gC}. When the system does not develop any order, i.e., $g_S=g_C=0$, then $\left\langle \mathcal{H}_i\rangle/\Lambda^2\right|_{\text{sol}}=0$. As we will show below, the system will always develop an order near the self-dual point.

For the mean field solution with a pure density wave order, one has
\begin{align} \label{eq:en-CDW}
\nonumber
&\left.\frac{\langle \mathcal{H}_i\rangle}{\Lambda^2}\right|_{\text{DW}}
\\ \nonumber
&=\left(\frac{\Delta}{\Delta-2}\right)\frac{\mathcal{E}}{\Lambda^2}
-\frac{1}{2z}\left(\frac{g_C^2/\Lambda^4}{\mathcal{J}_C/\Lambda^2}\right)
\\
&=\frac{K-2}{8}\left[\left(\frac{8}{4-K}\right)F\left(\frac{K-1}{K}\right)\right]^{\frac{4-K}{2-K}}
\left(\frac{z\mathcal{J}_C}{2\Lambda^2}\right)^{\frac{2}{2-K}}.
\end{align}
Therefore, the energy density is always negative when $K\approx 1$. This indicates that a formation of ordered state is energetically favorable in that region. At $K=1$, the density wave ordered state has
\begin{eqnarray} \label{eq:en-CDW-1st}
\left.\frac{\langle \mathcal{H}_i\rangle}{\Lambda^2}\right|_{\text{DW},\epsilon=0}
=-\left(\frac{\eta}{3}\right)^3
\left(\frac{4z\mathcal{J}_C}{\Lambda^2}\right)^2.
\end{eqnarray}
Similarly, one can calculate the corresponding expected energy density for a pure superfluid order,
\begin{align} \label{eq:en-SC}
\nonumber
&\left.\frac{\langle \mathcal{H}_i\rangle}{\Lambda^2}\right|_{\text{SF}}
\\
&=\frac{1-2K}{8K}\left[\left(\frac{8K}{4K-1}\right)F\left(1-K\right)\right]^{\frac{4K-1}{2K-1}}
\left(\frac{z\mathcal{J}_S}{2\Lambda^2}\right)^{\frac{2K}{2K-1}}.
\end{align}
At $K=1$, the superfluid ordered state has
\begin{eqnarray} \label{eq:en-SC-1st}
\left.\frac{\langle \mathcal{H}_i\rangle}{\Lambda^2}\right|_{\text{SF},\epsilon=0}
=-\left(\frac{\eta}{3}\right)^3
\left(\frac{4z\mathcal{J}_S}{\Lambda^2}\right)^2.
\end{eqnarray}
From the above results, one can conclude that the density wave order (superfluid order) is energetically favorable when $\mathcal{J}_C>\mathcal{J}_S$ ($\mathcal{J}_S>\mathcal{J}_C$) at $K=1$. This result agrees with the renormalization group analysis. At $K=1$, the interchain DW and SF coupling terms have the same scaling dimension. Hence, the one with larger bare coupling strength will dominate.

\subsection{Stability of the supersolid order near $K=1$}
\label{sec:stability-SS}

For the possible coexistence of DW and SF orders, the corresponding mean field solution for $\kappa_S=g_S/\Lambda^2$ and $\kappa_C=g_C/\Lambda^2$ are given by Eqs.~\eqref{eq:sol-ks2} and~\eqref{eq:sol-kc2}. Since both $\kappa_S$ and $\kappa_C$ are nonzero, the corresponding expectation value of the energy density is Eq.~\eqref{eq:exp-h}. In Appendix~\ref{app:technical}, we find that the following relationship holds,
\begin{eqnarray} \label{eq:relation}
-\frac{g_S^2}{z\mathcal{J}_S}
-\frac{g_C^2}{z\mathcal{J}_C}
=\frac{4(1-Z)}{3-4Z}\mathcal{E}.
\end{eqnarray}
Hence, one has
\begin{eqnarray} \label{eq:h-SS}
\left.\frac{\langle \mathcal{H}_i\rangle}{\Lambda^2}\right|_{\text{sol}}
=\left[\frac{\Delta}{\Delta-2}+\frac{2(1-Z)}{3-4Z}\right]\frac{\mathcal{E}}{\Lambda^2}.
\end{eqnarray}
Due to the complicated forms of $\mathcal{E}$, we are unable to obtain an analytic expression for $\left.\langle \mathcal{H}_i\rangle/\Lambda^2\right|_{\text{sol}}$ in the present case. Instead, we compare the energy density for supersolid solution with Eqs.~\eqref{eq:en-CDW} and \eqref{eq:en-SC} numerically. 

\begin{figure} [htb]
\includegraphics[width=2.25in]{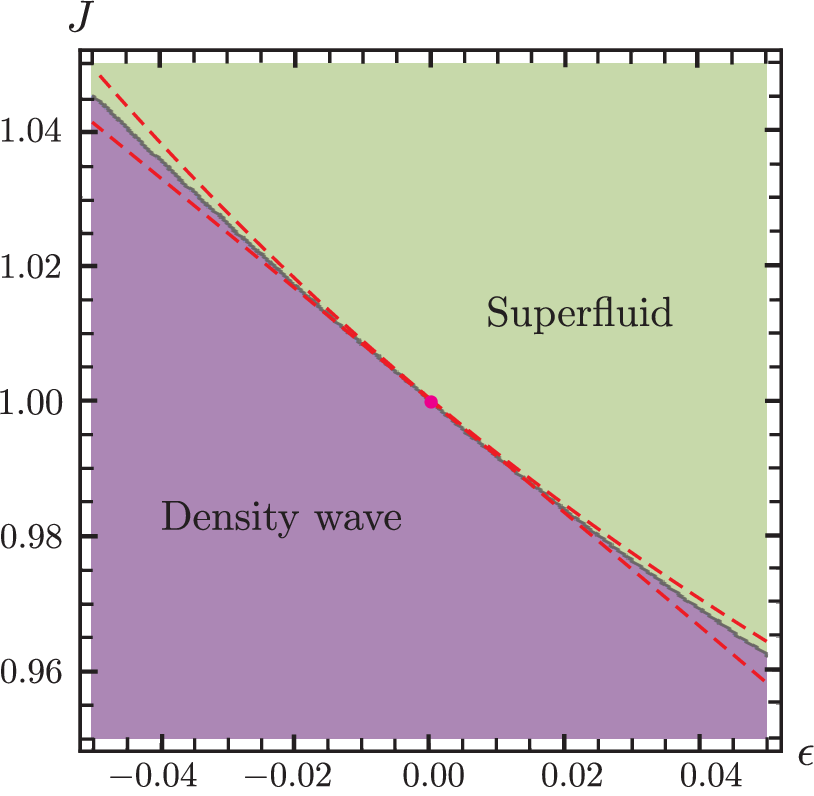} \\ \vspace{+0.2cm}
\includegraphics[width=2.25in]{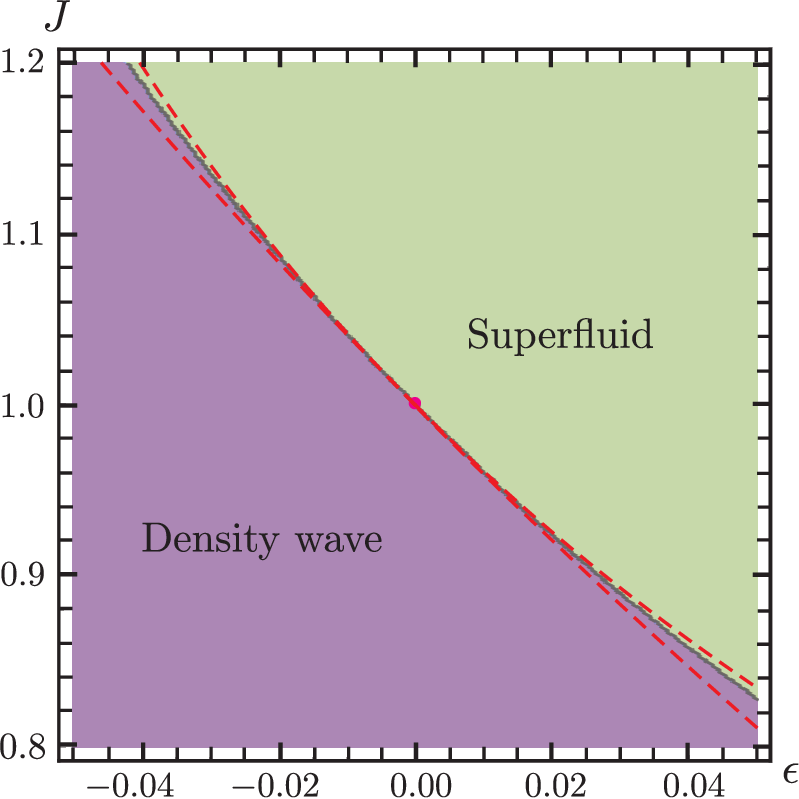} \\ \vspace{+0.2cm}
\includegraphics[width=2.25in]{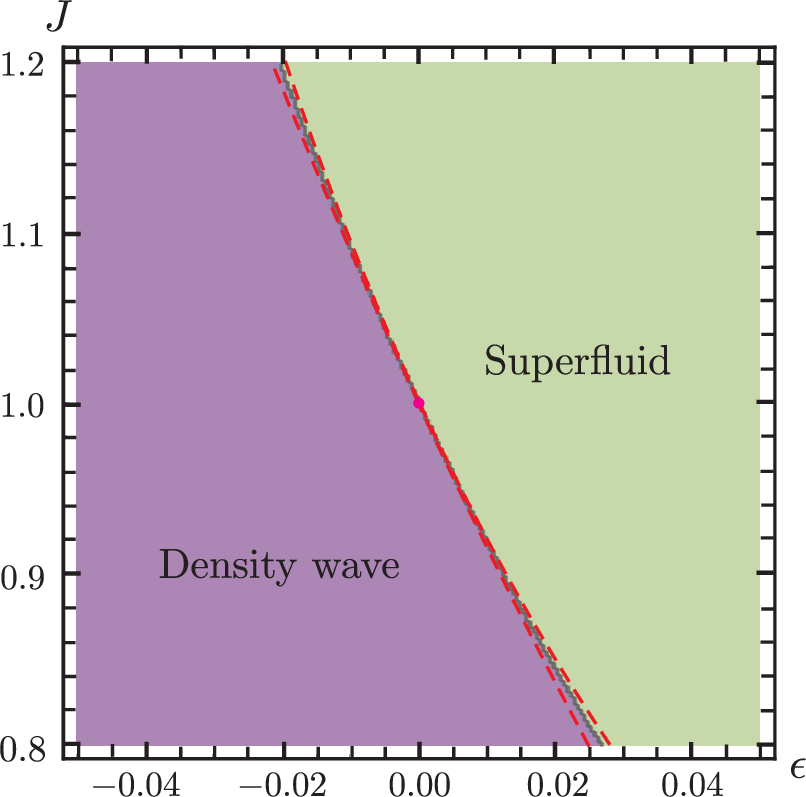}
\caption{Numerically predicted phase diagrams for the system near the self-dual point ($\epsilon=0$ and $J=\mathcal{J}_S/\mathcal{J}_C=1$), marked by the red dots. Here, the dimensionless effective interchain coupling strength $zj_S=z\mathcal{J}_S/\Lambda^2$ takes value $1/10$ (top panel), $1/50$ (middle panel), and $1/500$ (bottom panel). The region bounded by the red dashed lines is the region where the supersolid solution in Eq.~\eqref{eq:sol-ks2} and~\eqref{eq:sol-kc2} is well defined.}
\label{fig:phase-diagram}
\end{figure}

In Fig.~\ref{fig:phase-diagram}, we plot numerically the phase diagram of the system in the vicinity of the self-dual point for $j_S z=1/10$, $j_S z=1/50$, and $j_S z=1/500$ separately. To be specific, the ordered state that has the minimum value of $\left.\langle \mathcal{H}_i\rangle/\Lambda^2\right|_{\text{sol}}$ is shown at each point in the phase diagram. It is observed that either the density wave or the superfluid phase has the minimum energy density. At the self-dual point, the DW, SF, and the supersolid states have the same energy. In addition, the phase boundary between the DW and SF phases is located inside the region in which the supersolid order solution exists (bounded by the red dashed lines in the figure). A further numerical analysis (not shown here) verifies that $\left.\langle \mathcal{H}_i \rangle/\Lambda^2\right|_{\text{sol}}$ for the supersolid order solution is always larger than those for pure DW and SF order, when the system is sufficiently close to the self-dual point. From these features, we conclude that the solution for coexisting orders in Eqs.~\eqref{eq:sol-ks2} and~\eqref{eq:sol-kc2} corresponds to the scenario in Fig.~\ref{fig:landscape}(d), namely being a local maximum in the energy landscape. The absence of a region in which the coexisting phase is energetically favorable, and the position of the phase boundary suggest that the DW and SF phases are likely to be separated by a first-order phase transition. Finally, we should emphasize that our analysis considers the system at zero temperature only.

\section{Conclusion and discussion}
\label{sec:conclusion}

To summarize, we have revisited the possibility of realizing supersolidity for bosons trapped in an array of one-dimensional tubes, with coexistence of the density wave (DW) and superfluid (SF) orders. Following previous work~\cite{Carr}, we employ the mean field approximation to the interaction terms between neighboring tubes, and obtain a mean field Hamiltonian that takes the form of double sine-Gordon model. While this model is nonintegrable in general, it can be transformed into the integrable sine-Gordon model when the system has Luttinger parameter $K=1$. We took this special feature as our point of departure, and derived an approximate but integrable Hamiltonian that describes the system when it is near $K=1$. From this result, we have successfully obtained a set of approximate self-consistency equations for the DW and SF order parameters. This allowed us to study different ordered phases, and their competition or coexistence by comparing their corresponding energies quantitatively. Our results are asymptotically exact in the limits of $K-1\rightarrow 0$ and large dimensionality/coordination number.

We did find a mean field solution for coexisting density wave and superfluid order parameters [see Eqs.~\eqref{eq:sol-ks2} and~\eqref{eq:sol-kc2}], in a narrow region near the self-dual point. The region is determined by Eq.~\eqref{eq:sol-region}. As the interchain coupling or intertube coupling strengths become smaller relative to the UV energy cutoff, the region becomes narrower. 
We further performed a quantitative energy comparison between different solutions to the mean field equations. We found that the supersolid solution corresponds to a local maximum in the energy landscape of the system near the self-dual point. The analysis also shows that the critical line that separates the density wave and superfluid phases is well-located in the region in which the supersolid solution can be found. These observations lead us to the conclusion that the two phases are separated by a first order phase transition near the self-dual point. We emphasize that our work only focuses on the case with zero temperature. Without a suitable formulation of the Landau-Ginzburg free energy, we are unable to draw any firm conclusion for the phase diagram at nonzero temperature. It is possible that supersolidity becomes energetically favorable there. If that happens, it may serve as an example of the so-called ``order by disorder"~\cite{Balents2007}. We cannot rule out its presence far away from the self-dual point at zero temperature either.

Our results extend that of Ref.~\cite{Tsvelik2002}, which studied the competition between charge density wave and superconducting orders in high-$T_c$ superconducting cuprates, and include it as a special case. There, the interchain coupling terms were also treated by the mean field approximation. At $K=1$, this work concluded that coexisting CDW and SC orders can be achieved only when the two interchain coupling parameters have identical magnitudes, i.e. $\mathcal{J}_S=\mathcal{J}_C$. In other words, the solution only exists at the self-dual point. This feature has been reproduced in our present work. Importantly, our analysis and results go beyond the self-dual point. This is possible because we have successfully mapped the double sine-Gordon model (obtained after the interchain mean field approximation) to a standard sine-Gordon model for $K\approx 1$, by treating $K-1$ as an expansion parameter. A crucial step was to deduce how the spatial derivatives of the dual fields transform under a rotation in the order parameter space. We determined the transformation by comparing the operator product expansion between the fields before and after the rotation. This is somehow different from non-Abelian bosonization~\cite{Tsvelik-book}. There, the vertex operators and the spatial derivative of the field must have scaling dimensions one, so that they can be treated in a unified manner due to the su(2)$_1$ algebra they form.

On the other hand, Refs.~\cite{Fradkin2010, Carr} pointed out that the interchain mean field approximation may break down when the system goes beyond the quasi-one-dimensional regime and actually being two-dimensional. There, a nonlinear sigma model with an enhanced O(4) symmetry was formulated at the self-dual point. A further RG analysis of the model suggested that a coexistence between superfluid and density wave (or superconducting and charge density wave) orders can actually exist. Being argued as a marginally irrelevant term, no quantitative analysis for the symmetry breaking term that describes the departure from $K=1$ was performed. Since our approach (which is asymptotically exact in the high-dimensional limit) is different, we are unable to comment on the validity of that treatment, which appears to be specific to 2D. Furthermore, we should clarify again that our approach only applies to the vicinity of the self-dual point. Therefore, our work does not eliminate any possible coexistence of density wave and superfluid orders when the system is sufficiently far away from the self-dual point. 

We close by emphasizing our work has provided a quantitatively reliable treatment of the problem of competition between superfluid and density wave orders in the appropriate limits, and shed light on the broad issue of competing orders in strongly interacting systems.

\section*{Acknowledgment}

We thank Michael Levin for an inspiring discussion that turned out to be crucial for this work. The work of KKWM, OT and KY was performed at the National High Magnetic Field Laboratory, which is supported by National Science Foundation Cooperative Agreement No. DMR-1644779, and the State of Florida. KY's work is also National Science Foundation Grants No. DMR-1932796 and DMR-2315954. Work by AS is supported by the National Science Foundation under Grant No. DMR-2029401.

\onecolumngrid

\appendix

\section{Bosonization of the rotated XXZ model}
\label{app:bose-XXZ}

In the main text, we related the Luttinger liquid model that describes the coupled interchain systems to the XXZ model for quantum spins under a staggered magnetic field. Here, we will elaborate on this connection, perform the rotation in Eq.~\eqref{eq:transform} for the XXZ model, and present our argument on relating the resulting XXZ model to the Luttinger model. \\

We begin by restating the XXZ model with a staggered magnetic field in both $x$ and $z$ directions,
\begin{eqnarray} \label{eq:app-XXZ}
H_{XXZ}
=J \sum_j \left[\left(S_j^x S_{j+1}^x + S_j^y S_{j+1}^y\right)
+\Delta S_j^z S_{j+1}^z\right]
-h_x\sum_j (-1)^j S_j^x-h_z\sum_j (-1)^j S_j^z.
\end{eqnarray}
Recall the bosonization rules in Eqs.~(22) and~(23) in the main text~\cite{Cabra, Grynberg},
\begin{align} 
\label{eq:app-bos1}
S_z(x)
&=\frac{1}{\sqrt{2\pi}}\partial_x\phi(x)
+(-1)^j a_1\cos{\left[\sqrt{2\pi}\phi(x)\right]},
\\ 
\label{eq:app-bos2}
S_{\pm}(x)
&=e^{\pm i \sqrt{2\pi} \theta(x)}
\left\{ (-1)^j b_1 + b_2\cos{\left[\sqrt{2\pi}\phi(x)\right]}\right\}.
\end{align}
In writing the above equations, we have used the fact that the Fermi wavelength $k_F=\pi/2a$ when the spin system is half-filled. Here, the position of the spins are $x=ja$, where $a$ denotes the lattice spacing in the lattice. The constants $a_1$, $b_1$, $b_2$ are unimportant in our discussion. After bosonizing the XXZ model, one obtains
\begin{eqnarray} \label{eq:start-Luttinger}
H_L
=\int dx
\left\{
\frac{v}{2}
\left[K\left(\partial_x\theta\right)^2
+\frac{1}{K}\left(\partial_x\phi\right)^2\right]
-g_S \cos{(\sqrt{2\pi}\theta)}
-g_C \cos{(\sqrt{2\pi}\phi)}
\right\}.
\end{eqnarray}
Note that the Umklapp scattering term is neglected here since it is an ``artifact" of the half-filled spin system. It does not appear in the mean field Hamiltonian for our original problem of the coupled chain system. Eq.~\eqref{eq:start-Luttinger} is precisely the Hamiltonian that describes a single decoupled chain under the mean field approximation in the main text. When we bosonize the two terms for staggering field, only non-oscillatory terms are kept. Furthermore, we should recall that the Luttinger parameter $K$ \textit{cannot} be deduced from the above bosonization procedures (unless $\Delta\approx 0$ so that the result from perturbation theory holds). In general, the value of $K$ depends on $\Delta$, which can only be determined from the exact Bethe ansatz solution~\cite{Cabra},
\begin{eqnarray}
K=\frac{\pi}{\pi-\cos^{-1}{\Delta}}
\end{eqnarray}

In the main text, a rotation was performed for the Luttinger model at $K=1$ to transform the double sine-Gordon model into the standard sine-Gordon model. The corresponding transformation in $H_{XXZ}$ is a rotation of the spin quantization axis along the $y$-axis,
\begin{eqnarray} \label{eq:rotation-spin}
\begin{pmatrix}
\tilde{S}_j^x \\ \tilde{S}_j^z
\end{pmatrix}
=
\begin{pmatrix}
\cos{\alpha} & -\sin{\alpha} \\
\sin{\alpha} & \cos{\alpha}
\end{pmatrix}
\begin{pmatrix}
S_j^x \\ S_j^z
\end{pmatrix}.
\end{eqnarray}
We should emphasize that the rotation in $H_{XXZ}$ can be performed at any value of $\Delta$. On the other hand, the rotation in the Luttinger model can be performed only at $K=1$. Carrying out the rotation for $H_{XXZ}$ explicitly, we get
\begin{align} \label{eq:app-rotated-XXZ}
\nonumber
H'
=&~\sum_j 
\left[
\left(J_1'+J_2'\right)\tilde{S}_j^x \tilde{S}_{j+1}^x
+\left(J_1'-J_2'\right)\tilde{S}_j^y \tilde{S}_{j+1}^y
+J\left(\sin^2{\alpha}+\Delta\cos^2{\alpha}\right)\tilde{S}_j^z \tilde{S}_{j+1}^z
\right]
-\sqrt{h_x^2+h_z^2}\sum_j (-1)^j \tilde{S}_j^z
\\
&-J(\Delta-1)
\sin{\alpha}\cos{\alpha}
\sum_j
\left(
\tilde{S}_j^x \tilde{S}_{j+1}^z
+\tilde{S}_j^z \tilde{S}_{j+1}^x
\right).
\end{align}
Here, the parameters $\tan{\alpha}$, $J_1'$, and $J_2'$ are given by
\begin{align}
\tan{\alpha}
&=h_x/h_z
\\
J_1'
&=\frac{J}{2}\left(\cos^2{\alpha}+\Delta\sin^2{\alpha}+1\right)
\\
J_2'
&=\frac{J}{2}\left(\cos^2{\alpha}+\Delta\sin^2{\alpha}-1\right).
\end{align}
When $\Delta=1$ (i.e., the Heisenberg model or $K=1$ in the Luttinger model), it is obvious that $H'$ preserves the original SO(3) symmetry. Hence, we can immediately conclude that its bosonized form must take the same form as Eq.~\eqref{eq:start-Luttinger} at $K=1$ with the two cosine terms being combined. This explains Eq.~(24) in the main text. \\

Going away from $\Delta=1$ (i.e., $K\neq 1$ in the corresponding Luttinger model), $H'$ no longer takes the form of a XXZ model. As we only focus on the Luttinger model at $K\approx 1$, the corresponding $J_1'\approx J$ and $J_2'\approx 0$ in $H'$ as $\Delta\approx 1$. In other words, the first line of Eq.~\eqref{eq:app-rotated-XXZ} takes the form of an XYZ model with a small anisotropy $J_2'$ between the XY terms. This can be bosonized~\cite{Grynberg},
\begin{eqnarray}
H_1
=\int dx
\left\{
\frac{\tilde{v}}{2}
\left[\tilde{K} (\partial_x\tilde{\theta})^2
+\frac{1}{\tilde{K}} (\partial_x\tilde{\phi})^2\right]
-\sqrt{g_S^2+g_C^2} \cos{(\sqrt{2\pi}\tilde{\phi})}
+\lambda\cos{(2\sqrt{2\pi}\tilde{\theta})}
\right\}.
\end{eqnarray}
Here, $\lambda\propto J_2'\ll 1$. Notice that the Luttinger parameter $\tilde{K}\neq K$ in $H_L$. Meanwhile, we know that $\tilde{K}\approx 1$, so the term proportional to $\lambda$ is less relevant than the term which is proportional to $\sqrt{g_S^2+g_C^2}$. Now, we move to the bosonization for the second line of $H'$ and only keep the most relevant terms:
\begin{align}
H_2
=-\frac{J}{2}(\Delta-1)\sin{2\alpha}
\int dx~ \left\{
u_1\cos{(\sqrt{2\pi}\tilde{\theta})}
\cos{(\sqrt{2\pi}\tilde{\phi})}
+ u_2\cos{(\sqrt{2\pi}\tilde{\theta})}
\sin{(\sqrt{2\pi}\tilde{\phi})}
\right\}.
\end{align}
The bosonized model $H_1+H_2$ cannot be recast in the form of $H_L$ in Eq.~\eqref{eq:start-Luttinger}. This is expected as the original U(1) symmetry for the total magnetization in the XXZ model (here, we mean the model without the staggering field) has been hidden by the rotation in Eq.~\eqref{eq:rotation-spin}. When $\Delta\approx 1$, $H_2$ corresponds to a small nonlinear perturbation to $H_1$. Before the rotation, $H_{XXZ}$ should describe a gapless phase as the corresponding $H_L$ we considered in the main text does describe a gapless phase. This gaplessness cannot be altered by the rotation in the spin quantization axis. Hence, $H_2$ will need to flow to zero in the low-energy limit. This in turn renormalizes $\tilde{K}\rightarrow \tilde{K}'$ in $H_1$. Although we cannot determine the value of $\tilde{K}'$ by analyzing the XXZ model, the eventual fixed point of the continuum model must remain  a Luttinger liquid, with the Luttinger parameter being determined in the main text.

\section{Approximation to the region of supersolid solutions in Eq.~\eqref{eq:sol-region}}
\label{app:approx-region}

The region in which the supersolid solution exists (i.e. $g_S^2>0$ and $g_C^2>0$) are given in Eq.~\eqref{eq:sol-region}. Here, we derive an analytic approximation to the region. This provides a better understanding of the asymptotic behavior of the system near the self-dual point. Since $x=W_0(y)$ is the solution to the equation $y=xe^x$ and $W_0(y)$ is a monotonic increasing function for $y>0$, it is possible to invert Eq.~\eqref{eq:sol-region} for $\epsilon>0$ as follows,
\begin{eqnarray} \label{eq:region-appdenix}
\frac{3-\epsilon}{2+J}\exp{\left[\frac{(1-J)(3-\epsilon)}{2(2+J)\epsilon}\right]}
<\frac{3}{4z\eta\varsigma^{1/3}j_S}
<\frac{3+\epsilon}{1+2J}\exp{\left[\frac{(1-J)(3+\epsilon)}{2(1+2J)\epsilon}\right]}.
\end{eqnarray}
It is reminded that our work focuses on the region with $\epsilon\approx 0$ and $J=1$. Hence, we expand the inequality in the first order of $\epsilon$ and $J-1$. This gives
\begin{eqnarray}
\left(1-\frac{J-1}{3}-\frac{\epsilon}{3}\right) 
\exp{\left(-\frac{J-1}{2\epsilon}\right)}
\lesssim \frac{3}{4z\eta\varsigma^{1/3}j_S}
\lesssim \left[1-\frac{2(J-1)}{3}+\frac{\epsilon}{3}\right] 
\exp{\left(-\frac{J-1}{2\epsilon}\right)}.
\end{eqnarray}
At the first glance, there is an essential singularity in the exponential function. Nevertheless, we know that $J=1$ must be satisfied at $\epsilon=0$~\cite{Tsvelik2002}. Hence, the exponential function does not diverge there but being indeterminate. After a straightforward manipulation and only keeping the most significant terms, one can obtain
\begin{eqnarray} \label{eq:region1}
2\epsilon\left(1-\frac{2\epsilon}{3}\right)\ln{\left(\frac{4z\eta\varsigma^{1/3}j_S}{3}\right)}
-\frac{2\epsilon^2}{3}
\lesssim J-1
\lesssim
2\epsilon\left(1-\frac{4\epsilon}{3}\right)\ln{\left(\frac{4z\eta\varsigma^{1/3}j_S}{3}\right)}
+\frac{2\epsilon^2}{3}.
\end{eqnarray}
Using similar procedure, one can also obtain an approximate inequality for Eq.~\eqref{eq:sol-region} for $\epsilon<0$. It turns out that the result is the same as Eq.~\eqref{eq:region1}. It is reminded that 
$j_S=\mathcal{J}_S/\Lambda^2\ll 1$, so the logarithmic function is negative. Hence, the inequality is consistent with our assumption that $\epsilon$ and $J-1$ have opposite signs. Although $j_S\ll 1$, the logarithmic dependence of it strongly suppresses its contribution relative to the quadratic dependence of $\epsilon^2$. Therefore, the region in which the coexisting orders may exist becomes very narrow as $\epsilon\rightarrow 0$.

\section{Derivation of Eq.~\eqref{eq:relation}}
\label{app:technical}

In this appendix, we examine the self-consistency equations and derive a simple form for the energy of the mean field ground state in the supersolid state. First, we write Eq.~\eqref{eq:E_Z} in the following form
\begin{align}
\mathcal{E}(Z)
&=-\Lambda^2 F(Z)\left(\frac{g_S^2+g_C^2}{\Lambda^4}\right)^{2(1-Z)/(3-4Z)},
\\
Z
&=\epsilon\left(\frac{g_C^2-g_S^2}{g_C^2+g_S^2}\right)
=\epsilon\left(\frac{1-\mu^2}{1+\mu^2}\right).
\end{align}
Without expanding $\mathcal{E}(Z)$ in a power series of $\epsilon$, the self-consistency equations are
\begin{align}
-\frac{g_S}{z\mathcal{J}_S}
&=\frac{\mathcal{E}}{g_C}
\left\{
\frac{\partial\ln{F}}{\partial\mu}
+\frac{\partial}{\partial Z}\left(\frac{2-2Z}{3-4Z}\right)\frac{\partial Z}{\partial\mu}
\ln{\left(\frac{g_S^2+g_C^2}{\Lambda^4}\right)}
+\frac{4(1-Z)}{3-4Z}\left(\frac{g_C}{g_S}\right)
\left(\frac{g_S^2}{g_S^2+g_C^2}\right)
\right\},
\\
-\frac{g_C}{z\mathcal{J}_C}
&=-\frac{g_S \mathcal{E}}{g_C^2}
\left\{
\frac{\partial\ln{F}}{\partial\mu}
+\frac{\partial}{\partial Z}\left(\frac{2-2Z}{3-4Z}\right)\frac{\partial Z}{\partial\mu}
\ln{\left(\frac{g_S^2+g_C^2}{\Lambda^4}\right)}
-\frac{4(1-Z)}{3-4Z}\left(\frac{g_C}{g_S}\right)
\left(\frac{g_C^2}{g_S^2+g_C^2}\right)
\right\}.
\end{align}
By a simple rearrangement of terms, one has
\begin{align}
-\frac{g_C}{g_S}\left(\frac{g_S^2}{z\mathcal{J}_S \mathcal{E}}\right)
&=
\frac{\partial\ln{F}}{\partial\mu}
+\frac{\partial}{\partial Z}\left(\frac{2-2Z}{3-4Z}\right)\frac{\partial Z}{\partial\mu}
\ln{\left(\frac{g_S^2+g_C^2}{\Lambda^4}\right)}
+\frac{4(1-Z)}{3-4Z}\frac{g_C}{g_S}\frac{g_S^2}{g_S^2+g_C^2},
\\
\frac{g_C}{g_S}\left(\frac{g_C^2}{z\mathcal{J}_C\mathcal{E}}\right)
&=
\frac{\partial\ln{F}}{\partial\mu}
+\frac{\partial}{\partial Z}\left(\frac{2-2Z}{3-4Z}\right)\frac{\partial Z}{\partial\mu}
\ln{\left(\frac{g_S^2+g_C^2}{\Lambda^4}\right)}
-\frac{4(1-Z)}{3-4Z}\frac{g_C}{g_S}\frac{g_C^2}{g_S^2+g_C^2}.
\end{align}
From the above two equations, we immediately obtain 
\begin{eqnarray}
-\frac{g_S^2}{z\mathcal{J}_S}-\frac{g_C^2}{z\mathcal{J}_C}
=\frac{4(1-Z)}{3-4Z}\mathcal{E}.
\end{eqnarray}
This is exactly Eq.~\eqref{eq:relation} in the main text.

\twocolumngrid


\begin{thebibliography}{99}
\bibitem{footnote1}
Note that the corresponding regime is $1/4<K<1$ in other references with a different normalization convention~\cite{Carr}).
\bibitem{footnote-Ukmlapp}
In fact, the possible Ukmlapp process in the spin system will lead to an additional cosine term after bosonization, $\int dx~\lambda\cos{(2\sqrt{2\pi}\phi)}$. However, this term is specific to the half-filled system, and does not affect our discussion in the main text.
\bibitem{footnote-productlog}
The Lambert $W$ function is also known as the product logarithm function. For $y>0$, the equation 
$xe^x=y$ is completely solved by $x=W_0(y)$. Here, $W_0(y)$ stands for the principal branch of the Lambert $W$ function. If $-1/e<y<0$, then there are two different solutions to the equation. Besides $x=W_0(y)$, another possible solution is given by $x=W_{-1}(y)$.
\end{thebibliography}
\end{document}